\documentclass[twocolumn,floatfix,showpacs,eqsecnum,prb,superscriptaddress]{revtex4}
\usepackage{graphicx}
\usepackage{graphics}
\usepackage{amsmath}
\usepackage{amssymb}
\usepackage{bm}
\usepackage{color}
\usepackage[normalem]{ulem}

\newcommand{\ket}[1]{|{#1}\rangle}
\newcommand{\bra}[1]{\langle{#1}|}

\newcommand{\FFF}{\mathcal{F}}

\begin{document}

\title{Scattering theory of topological phases in discrete-time quantum walks}

\author{B. Tarasinski}
\affiliation{Instituut-Lorentz, Universiteit Leiden, 
P.O. Box 9506, 2300 RA Leiden, The Netherlands}
\author{J.~K.~Asb\'oth}
\affiliation{Institute for Solid State Physics and Optics, 
Wigner Research Centre, Hungarian Academy of Sciences, 
H-1525 Budapest P.O. Box 49, Hungary}
\author{J. P. Dahlhaus}
\affiliation{Department of Physics, University of California, 
Berkeley, California 95720, USA}
\date{January 2014}

\begin{abstract}

One-dimensional discrete-time quantum walks show a rich spectrum of
topological phases that have so far been exclusively analysed
based on the Floquet operator in momentum space. In this work
we introduce an alternative approach to topology which is based on the
scattering matrix of a quantum walk, adapting concepts from
time-independent systems.  For quantum
walks with gaps in the quasienergy spectrum at $0$ and $\pi$, we find three different types of topological invariants, which apply dependent on the symmetries of the system. These determine the number of protected boundary states at an interface between two quantum walk regions. 
Unbalanced quantum
walks on the other hand are characterised by the number of
perfectly transmitting unidirectional modes they support, which is equal to their non-trivial quasienergy winding. Our classification provides a unified
framework that includes all known types of topology in one dimensional
discrete-time quantum walks and is very well suited for the analysis
of finite size and disorder effects. We provide a simple scheme to
directly measure the topological invariants in an optical quantum walk
experiment.
\end{abstract}
\pacs{03.65.Vf, 42.50.-p, 05.30.Rt}
\maketitle

\section{Introduction}

The last decade has seen a systematic exploration of
topological phases in band insulators and the protected low energy
states that emerge at their boundaries.\cite{rmp_kane,rmp_zhang} From Majorana
bound states at the ends of topological superconducting wires to 
the unique metallic surface state of three-dimensional topological
insulators, a variety of boundary states can arise 
in this way. Their
potential applications range from spintronics to topological quantum
computation. As there are few real-life materials that are topological
insulators,\cite{ando_materials2013} there is an intense search for
model systems that simulate topological insulators in the
laboratory.\cite{hofstadter_ultracold2013,kraus_fibonacci2012,
  sun_topological_2012}

Discrete-time quantum walks (DTQW)\cite{
venegas_2012} are quantum
generalizations of the random walk, with a quantum speedup that could
be employed for fast quantum search \cite{qw_search_2002} or even for
general quantum computation.\cite{dtqw_universal} They have been
realized in many experimental setups, including atoms in optical
lattices,\cite{meschede_science,alberti_electric_experiment} trapped
ions,\cite{schmitz_ion,roos_ions} and light in optical setups.
\cite{peruzzo_science_2010, white_photon_prl, schreiber_science,
  gabris_prl, sciarrino_twoparticle, delayed_choice_experiment}
DTQWs are known to simulate topological
  insulators,\cite{kitagawa_exploring} this was
  recently experimentally confirmed by the observation of edge
states in an inhomogeneous quantum walk with
photons. \cite{kitagawa_observation}
  
Beyond realising entries in the periodic table of
  topological insulators,\cite{schnyder_tenfold} DTQWs possess a
  richer structure of topological phases which is subject of ongoing
  research.  The role of energy is taken over by quasienergy
$\varepsilon$, that is $2\pi$-periodic in natural units, where
$\hbar=1$ and the unit of time is one timestep of the walk. This is a
feature that quantum walks share with periodically driven
lattice Hamiltonians,\cite{floquet_topological,dora_review} for which
unique topological invariants have been found.
\cite{kitagawa_periodic}  For both types of systems, topologically
protected states may appear both at quasienergy $\varepsilon=0$ and
$\varepsilon=\pi$,\cite{akhmerov_majorana_driven} and states may be
topologically protected even when all bands are topologically trivial.
\cite{rudner_driven,kitagawa_introduction}

In this work we characterize topological phases of one dimensional
DTQWs using a scattering matrix approach. This constitutes a
generalization of methods developed for time-independent systems.\cite{akhmerov_scattering_2011,fulga_scattering_2011,
  dahlhaus_scattering_2012} For DTQWs
with  gaps in the
quasienergy spectrum at both $\varepsilon=0$ and $\varepsilon=\pi$, 
we obtain the topological invariants as simple functions of the
scattering matrix at these quasienergies. 
For unbalanced quantum walks, where there is an unequal number of
left- and rightward shifts in a period, we find an integer number of
perfectly transmitting unidirectional modes, that is equal to the
quasienergy winding.\cite{kitagawa_periodic} Our approach is
particularly suitable to calculate the topological invariants of
disordered quantum walks, as we demonstrate in an example.

This paper is structured as follows. After defining our notation for
one-dimensional discrete-time quantum walks in the next section, we
adapt the concept of a scattering matrix for DTQWs in
Sec.~\ref{sect:scattering}.  In Sec.~\ref{sect:symmetries} we discuss
the influence of particle-hole, time-reversal and chiral symmetry on
the scattering matrix.  The central result of our paper, the
topological invariants of DTQWs, are shown in Sections
\ref{sect:topological_gapped} and \ref{sect:topology_winding}.  We
illustrate our approach in Sec.~\ref{sec:Examples} with concrete
examples.  Finally, Sec.~\ref{sect:experiment} discusses how the topological invariants can be
directly measured in a quantum walk experiment.

\section{Discrete-time quantum walks}
\label{sect:notation}

We consider a particle (walker) with $N$ internal states (coin
states) on a one-dimensional lattice, whose wave function can be
written as
\begin{align}
\ket{\Psi} &= \sum_{x\in \mathbb{Z}} \sum_{n=1}^{N} \Psi(x,n)
\ket{x,n}.
\label{eq:def_basis} 
\end{align}
Here $x$ denotes the discrete position and $n$ the internal state of
the walker.

The walker is subjected to a periodic sequence of two different types
of operations: shifts and rotations. Measuring time $\tau$ in units of
the period, the dynamics are given by
\begin{align}
\ket{\Psi(\tau+1)} &= \mathcal{F}\ket{\Psi(\tau)},\\
\mathcal{F}&=R_{M+1}S_{M} R_{M}\ldots S_1 R_1.
\label{eq:evolution}
\end{align}
The time-evolution operator over one period, a.k.a. Floquet operator
$\mathcal{F}$, consists of shift operators $S_{j}$ and rotation
operators $R_{j}$. 

Each shift operation $S_j$, shifts a chosen internal state $n_j$ by
one lattice site, either to the right ($+$) or to the left ($-$). In
formulas $S_j=S^{\pm}_{n_j}$, with
\begin{align}
S^{\pm}_n &= \sum_{x\in \mathbb{Z}} \Big[ \ket{x\pm 1,n} \bra{x,n}+\!\!\sum_{n'\neq n}\!  \ket{x,n'} \bra{x,n'}\Big].
\label{eq:shift_def}
\end{align}
For each internal state $n$, we fix a direction $s_n \in \{+1,-1,0\}$
throughout the protocol. We require that the operators $S_j$ are
compatible with each other, i.e. no state is shifted to the left by
some $S_j$ and to the right by others. Accordingly, there are three
sets of internal states: those shifted to the right, $n \in M_+$,
those shifted to the left, $n \in M_-$, and those not shifted at all,
$n \in M_0$. For each internal state $n$, we use $d_n$ to denote the
number of shift operators $S$ in a period that shift the state,
\begin{align}
d_n &= \sum_{j=1}^{M} \delta_{n_j,n}.
\end{align}

Rotations mix the internal degrees of freedom, but are local in real
space,
\begin{align}
R_j &= \sum_{x\in \mathbb{Z}} \ket{x}\bra{x} \otimes R_j(x). 
\end{align}
Each $R_j(x)$ is a $U(N)$ operation. For translation invariant
quantum walks, $R_j(x)$ is independent of $x$. 

The time evolution $(\ref{eq:evolution})$ is a stroboscopic simulation of an
effective, time-independent Hamiltonian 
\begin{align}
H_{\rm eff}\equiv i \log \mathcal{F}.
\label{eq:effectiveH}
\end{align}
For definiteness, the branch cut of the logarithm is chosen such that all
quasienergies, the eigenvalues of $H_{\rm eff}$, are restricted to $\varepsilon
\in [-\pi,\pi)$. In the presence of translational symmetry, quantum walks thus have
 a band structure, just like time-independent systems. 
 
\begin{figure}[tb] 
\centerline{\includegraphics[width=1.0\linewidth]{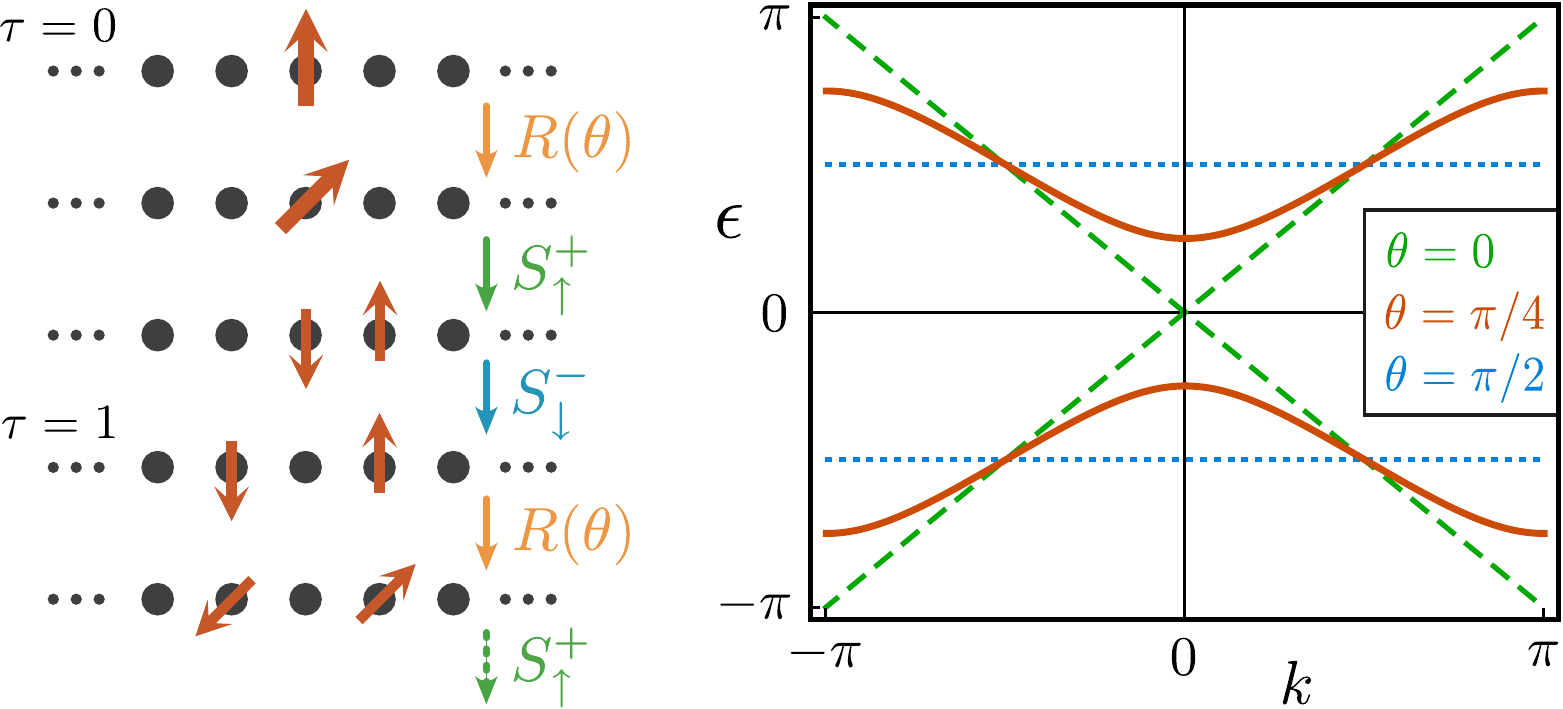}}
\caption{Left: Propagation of a particle in the simple quantum walk,
  initialized in spin-up state on a single site. Right: band structure
  of the simple quantum walk for different values of the rotation
  angle $\theta$. Generically the spectrum is gapped around quasienergies
  $\epsilon=0,\pi$ except for the special cases
  $\theta=0,\pi$.}
\label{fig:standardQW}
\end{figure}

 As an example, Fig. \ref{fig:standardQW} illustrates the protocol and
 the quasienergy band structure of the 
 simple quantum walk,
\begin{align}
\mathcal{F}=S^-_{\downarrow} S^+_{\uparrow} R(\theta).
\label{eq:simplQW}
\end{align}
The walker here has only two internal states, which we label by
  $\uparrow$ for $n=1$ and $\downarrow$ for $n=2$, and refer to as
spin. First the spinor is rotated by an angle $\theta$ on the Bloch
sphere,
\begin{align}
R(\theta)= \sum_x |x\rangle\langle x| \otimes e^{-i\theta \sigma_y}.
\label{eq:real-spin-rotation-matrix}
\end{align}
Subsequently
$S^+_\uparrow$ shifts the spin-up component of the state to the right
and $S^-_\downarrow$ the spin-down component to the left.

Note that the Floquet operator is not unique for a given quantum walk protocol. For example we could just as well choose 
\begin{align}
\mathcal{F}=S^+_{\uparrow} R(\theta)S^-_{\downarrow},
\label{eq:simplQW2}
\end{align}
for the Floquet operator of the simple quantum walk, since it produces the same protocol of operations ($\ldots S^+_{\uparrow} R(\theta)S^-_{\downarrow} S^+_{\uparrow} R(\theta)S^-_{\downarrow} \ldots$). Describing a quantum walk by a specific Floquet operator amounts to fixing a starting time, or time frame,\cite{asboth_2013} for the period of the walk. Changing the starting time of the period is much like
choosing a different unit cell in a crystal. It corresponds to a
unitary tranformation on the Floquet operator $\FFF$, and, as a
result, cannot change the quasienergy spectrum. Nevertheless, the choice of the correct time frame can be crucial when investigating symmetries and topological properties as we shall discuss in the course of the paper.

\section{Scattering in quantum walks}
\label{sect:scattering}

\begin{figure}[tb] 
\centerline{\includegraphics[width=0.9\linewidth]{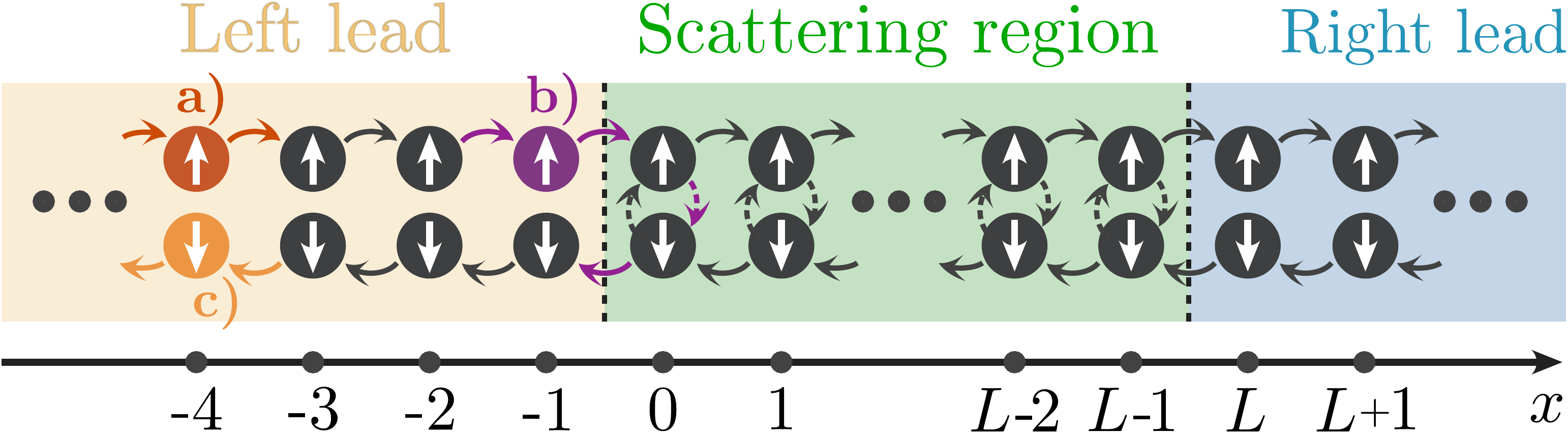}}
\caption{Scattering setting for the simple quantum walk,
  Eq.~\eqref{eq:simplQW}. The lattice is divided in three regions: a left lead ($x<0$), a right lead ($x\ge L$) and a scattering region in   
  between. Each site contains two internal spin states. 
  The shift operators of the protocol act throughout the whole system (solid black arrows), shifting a walker
  with state $\uparrow$ to the right, and state $\downarrow$ to the
  left. Rotations (dotted arrows) only change the
  internal state of the walker in the scattering region. \textbf{a)} A walker with spin-up in the left lead is incident on the scattering region. \textbf{b)} Once it reaches $x=0$, it is subject to rotations and acquires a spin-down component, which is shifted in the opposite direction. The purple arrows illustrate a possible reflection process. \textbf{c)} A walker with spin-down is propagated away from the scattering region. While \textbf{a)}-\textbf{c)} depict the scattering in time, the scattering states we consider are the corresponding quasienergy eigenstates.
}
\label{fig:scattering}
\end{figure}

To study DTQWs in a scattering setting, we maintain the whole quantum
walk protocol only in a central region ($0\le x <L$), which we want to
analyse. 
In the remaining regions we omit the rotations,
\begin{align}
R_j(x<0) &= R_j(x \ge L)=\openone_N \hspace{10pt} {\rm for\; all \;}j.
\end{align}
In this way, a left ($x< 0$) and a right lead ($x\ge L$) are formed. The scattering setting is illustrated in Fig.~\ref{fig:scattering} for the example of the simple quantum walk. Deep in the leads, a particle with internal state $n$ is simply shifted by $d_n$
sites in direction $s_n$ in each period,
\begin{align}
\mathcal{F}\ket{x,n}&=S_M\ldots S_1\ket{x,n}=\ket{x+s_n d_n,n}, \nonumber\\
\text{for  } & \quad x < -d_n 
\quad \text{ or  } 
\quad x > L + d_n.
\end{align}
An infinite lead of this type has propagating solutions at all quasienergies.

A natural basis for propagating states in the two leads ($l$,$r$) is
given by the states
\begin{align}
\ket{l_{n,d,\varepsilon}} &= \sum_{j=-\infty}^{0} e^{ i s_n \varepsilon j
} \ket{ j d_n- d, n}, \nonumber\\
\ket{r_{n,d,\varepsilon}}  &= \sum_{j=1}^\infty e^{i s_n\varepsilon j} 
\ket{L + j d_n- d,n},
\label{eq:lead_plane_waves}
\end{align}
for $n \in M_+\cup M_-$.  These are quantum walk equivalents of
plane waves, restricted to the left/right lead and normalized to carry the same particle current. Unlike true plane waves,\cite{meyer} they only occupy every $d_n$th site and the different sublattices that arise in this way are indexed
by $d$, restricted to $1\le d \le d_n$.

In a scattering problem, an incoming mode incident on a central region
is scattered into outgoing modes.  Consider a mode
$\ket{l_{n,d,\varepsilon}}$ in the left lead, with $s_n=+1$, so that
it is incident on the central region. It is scattered into outgoing
modes $\ket{\Psi_{L,R}^{\rm out}}$ in both the left and the right
lead.  The corresponding scattering state is a Floquet eigenstate with
quasienergy $\varepsilon$,
\begin{align}
\ket{\Psi_{n,d,\varepsilon}} =&\;
\ket{l_{n,d,\varepsilon}}+\ket{\Psi_C}+\ket{\Psi_L^{\rm out}}
+\ket{\Psi_R^{\rm out}},\\
\ket{\Psi_L^{\rm out}}=& \sum_{n'\in M_-} \sum_{d'} r_{n'd',nd}(\varepsilon)\,  
\ket{l_{n',d',\varepsilon}},\\
\ket{\Psi_R^{\rm out}}=& \sum_{n'\in M_+} \sum_{d'} t_{n'd',nd}(\varepsilon)\,
\ket{r_{n',d',\varepsilon}},
\label{eq:scattering-state-wave-function}
\end{align}
where $\ket{\Psi_C}$ denotes the contribution of the state in the
central region. This defines the matrix elements of both the reflection
matrix $r(\varepsilon)$ and the transmission matrix $t(\varepsilon)$.

Using the Floquet operator of the scattering setting, we can write
down the scattering state explicitly,
\begin{align} 
\ket{\Psi_{n,d,\varepsilon}} &= \sum_{\nu=-\infty}^{\infty} e^{i\varepsilon \nu}
\FFF^\nu \ket{-d,n}. 
\label{eq:scat_state}
\end{align} 
This really is a stationary state with quasienergy $\varepsilon$, as
can be seen by application of $\FFF$ on Eq.~
\eqref{eq:scat_state}. State $\ket{\Psi_{n,d,\varepsilon}}$ contains
the correct incoming plane wave, since
\begin{align} 
\ket{l_{n,d,\varepsilon}}= \sum_{\nu=-\infty}^{0} e^{i\varepsilon \nu}
\FFF^\nu \ket{-d,n}.
\label{eq:inc_state}
\end{align}
Furthermore, this state contains no incoming plane waves other than
$\ket{l_{n,d,\varepsilon}}$, since terms in the above sum with $\nu>0$
correspond to states that can be reached by propagating $\ket{-d,n}$
forward in time: they are in the central region and in the outgoing
modes.

The reflection matrix elements are found from projections of
$\ket{\Psi_{n,d,\varepsilon}}$ onto outgoing ($s_{n'}=-1$) states in
the left lead, $\ket{l_{n',d',\varepsilon}}$. Using 
the definitions above, we obtain 
\begin{align}
r_{n'd',nd}(\varepsilon) &=
\bra{-d',n'} \sum_{\nu=-\infty}^{\infty} e^{i\varepsilon \nu}\FFF^{\nu}
\ket{-d,n} \nonumber \\
\quad &=
 \bra{-d',n'} (1- e^{i\varepsilon }\FFF )^{-1} \ket{-d,n}.
\label{eq:scat_overlap}
\end{align}
Similarly, the transmission matrix elements are 
\begin{align}
t_{n'd',nd}(\varepsilon) &=     
 \bra{L-d',n'} (1- e^{i\varepsilon }\FFF )^{-1} \ket{-d,n}.
\label{eq:scat_overlap_t}
\end{align}
for all $n'$ with $s_{n'}=+$. For numerical evaluation, the reflection
and transmission matrices can be calculated from this formula using
Floquet operators that are truncated in the leads. We discuss this in
detail in Appendix~\ref{sec:numerical-implementation}.

Scattering matrices for DTQWs have been considered in a
  different formalism by Feldman and Hillery.\cite{feldman_hillery_1,
    feldman_hillery_2} With an elegant mathematical duality
  transformation, they assign the walker to the edges rather than the
  nodes. We chose a different route from theirs, as outlined in this
  Section, for two reasons. First, our approach is easier to apply to
  multistep walks (i.e., DTQWs where the number of steps per cycle is
  $M>2$). Second, and this is the more important reason: 
our approach allows for a transparent treatment of the relevant symmetries 
of the system. This is the topic we turn to in the next Section.

\section{Symmetries of quantum walks}
\label{sect:symmetries}

The standard band theory of topological insulators describes
topological phases of Hamiltonians depending on three discrete
symmetries: time-reversal symmetry (TRS), particle-hole symmetry
(PHS), and chiral symmetry (CS). In this section we show how the
definition of these symmetries translates to the Floquet operator and
the scattering matrix of DTQWs.


A quantum walk has TRS if an antiunitary operator
$\mathcal{T}=K U_T$ exists such that
\begin{align}
 U_T^{\dagger} \FFF^\ast U_T = \FFF^{-1} \quad & \Leftrightarrow 
\quad  U_T^{\dagger} H_\mathrm{eff}^\ast U_T = H_\mathrm{eff}. 
\end{align}
Here $K$ denotes complex conjugation in the basis used in
Eq.~\eqref{eq:def_basis}, and $U_T$ is a unitary operator acting on the internal state
only. The TRS operator $\mathcal{T}$ transforms the time-evolution
operator $\FFF$ into its inverse, justifying the term
``time-reversal''.

If a unitary operator $\Gamma$ achieves time reversal, this is
referred to as CS,
\begin{align}
 \Gamma^{\dagger} \FFF \Gamma = \FFF^{-1} \quad & \Leftrightarrow 
\quad  \Gamma^{\dagger} H_\mathrm{eff} \Gamma = -H_\mathrm{eff}. 
\end{align}

Finally, consider an anti-unitary operator $\mathcal{P}=K U_P$ that
transforms the Floquet operator into itself,
\begin{align}
 U_P^{\dagger} \FFF^\ast U_P = \FFF \quad & \Leftrightarrow 
\quad  U_P^{\dagger} H_\mathrm{eff}^\ast U_P = -H_\mathrm{eff}. 
\end{align}
A symmetry of this form is referred to as PHS, because of its
existence in superconductors. In quantum walks, there is no natural
concept of particles and holes, but a symmetry of this form might
still be present.

Like in the symmetry classification of time-independent problems, the
unitary symmetries present in the system are used to block diagonalize
the Floquet operator (and, as a consequence, the effective
Hamiltonian) before PHS, TRS and CS are analysed.  Then, $\mathcal{P}$
and $\mathcal{T}$, if present, will square to plus or minus unity,
and chiral symmetry is related to the two by $\Gamma \propto
\mathcal{TP}$, if both are present. The possible presence or absence,
as well as the squares of these symmetries, gives ten possible
symmetry classes, which are referred to by so-called Cartan
labels.\cite{altland_zirnbauer_1997,schnyder_tenfold}

\begin{table*}
\centering
\begin{tabular}{ | l || c | c | c | c | c |}
\hline
Symmetry class\ &  AIII &  CII  & BDI & D & DIII\\ \hline \hline
$\mathcal{T}^2$ & $\times$ & $-1$ &  $+1$ & $\times$ & $-1$\\ \hline
$\mathcal{P}^2$ & $\times$ &  $-1$ & $+1$ & $+1$ & $+1$  \\ \hline
$\Gamma$    & \checkmark & \checkmark &\checkmark & $\times$ & \checkmark  \\ 
\hline\hline
$\mathcal{Q}_X$=$\mathcal{Q}_{X,0}\times\mathcal{Q}_{X,\pi}$& 
$\mathbb{Z} \times\mathbb{Z}$ & 
$\mathbb{Z} \times\mathbb{Z}$ &
$\mathbb{Z} \times\mathbb{Z}$ & 
$\mathbb{Z}_2 \times\mathbb{Z}_2$ & 
$\mathbb{Z}_2 \times\mathbb{Z}_2$  \\ 
\hline
 & $\tfrac{1}{2}\text{Tr}\,r(0) \times 
\tfrac{1}{2}\text{Tr}\,r(\pi)$ & 
$\tfrac{1}{2}\text{Tr}\,r(0) \times 
\tfrac{1}{2}\text{Tr}\,r(\pi)$ &
$\tfrac{1}{2}\text{Tr}\,r(0) \times 
\tfrac{1}{2}\text{Tr}\,r(\pi)$ & 
$\tfrac{1}{2}\text{Det}\,r(0) \times
       \tfrac{1}{2} \text{Det}\,r(\pi)$ & $\text{Pf}\,r(0) \times \text{Pf}\,r(\pi)$  \\ \hline
\end{tabular}
\caption{Symmetry classes with non-trivial topological invariants in
  gapped one-dimensional DTQWs. For TRS and PHS, the table gives the
  square values of the symmetry operators. For CS, existence is
  indicated by $\checkmark$. The full topological invariant
  $\mathcal{Q}_X$ is composed of invariants $Q_{X,\varepsilon}$ at
  quasienergies $\varepsilon=0,\pi$ inside the two gaps of the
  quasienergy spectrum. The invariants as given in the table apply after a basis change on the reflection matrix, as detailed
  in Appendix~\ref{sec:basis-transformation-appendix}. }
\label{table}
\end{table*}

We now turn to the discussion of symmetries in a scattering setup. The
situation is is very similar to systems whose dynamics are governed by
time-independent Hamiltonians. We thus refer the reader especially to
Appendix~A of Ref.~\onlinecite{fulga_scattering_2012}.

If a scattering setup possesses one of the symmetries above, we can
consider the action of the symmetry operators on the modes in the leads.  TRS and CS reverse the action of the
time evolution operator, and thus map  incoming modes to outgoing
modes and vice versa, while PHS will act on these spaces separately.

We thus can write a time-reversed incoming state as a
superposition of outgoing states. In the left lead this reads:
\begin{align}
    \mathcal{T}\ket{l_{n,d,\epsilon}}  & = \sum_{n' \in M_-} Q_{T,n'n} 
\ket{l_{n', d, \epsilon}} \text{ for } n \in M_+.
    \label{eq:action-of-trs-on-scattering}
\end{align}
In the same manner, time-reversed outgoing states are superpositions of incoming states, with coefficients captured in the left lead by a matrix $V_{T}=\mathcal{T}^2 (Q_T)^T$.  Similarly, the action of CS is given by matrices
$Q_\Gamma$ and $V_\Gamma=\Gamma^2 Q_\Gamma^\dag$.  PHS on the other hand acts on right and
left moving states separately, and we write
\begin{align}
    \mathcal{P}\ket{l_{n, d, \epsilon}} = \sum_{n' \in M_\pm} Q_{P\pm,n'n} 
\ket{l_{n', d, \epsilon}} \text{ for } n \in M_\pm.
    \label{eq:phs-action-on-lead-states}
\end{align}
Here the matrices $V_{P+}$ and $V_{P-}$ are independent and, in
general, can have different dimensions. 

The symmetries of the Floquet operator $\FFF$ translate to properties
of the reflection matrix $r$:
\begin{align} 
    r(\varepsilon) &= Q_T r(\varepsilon)^T V_T^\dag, 
\label{eq:trs-of-reflection-block}\\
    r(\varepsilon) &= Q_\Gamma r(-\varepsilon)^\dag V_\Gamma^\dag, 
\label{eq:cs-of-reflection-block}\\
	r(\varepsilon) &= Q_{P-} r(-\varepsilon)^* Q_{P+}^\dag.  
\label{eq:phs-of-reflection-block}
\end{align}

There is an important caveat here. The Floquet operator, and,
consequently, the effective Hamiltonian and the scattering matrix, all
depend on the choice of time frame, as in the example of
Eq.~\eqref{eq:simplQW2}. As a consequence, the same DTQW can be seen
to have a symmetry in one timeframe, while this symmetry might be
hidden in another timeframe -- this holds especially for TRS and CS.
 Therefore, finding the symmetries and the topological
invariants includes going into the proper timeframe. In this Section
and in the rest of the paper, we assume that this work has been done
and that we are in a timeframe where the symmetries are explicit.

There are two special quasienergies for a DTQW. 
As seen from
Eqs. \eqref{eq:cs-of-reflection-block} and
\eqref{eq:phs-of-reflection-block}, CS and PHS yield special
constraints on the scattering matrix if $\varepsilon=-\varepsilon$,
which, due to the periodicity of quasienergy, is fulfilled at both
$\varepsilon=0$ and $\varepsilon=\pi$.  As we show in the following,
this has the consequence that for DTQWs, topological invariants come
in pairs.

\section{Topological invariants of gapped quantum walks}
\label{sect:topological_gapped}

In this section we consider balanced quantum walks, where the number $n_+$ of shift operators that shift to the
    right
equals the number $n_-$ of shift operators that shift
  to the left in a period.
For these walks, the quasienergy band structure 
generically has 
gaps around the special quasienergies $\varepsilon=0$ and
$\varepsilon=\pi$. Then, the transmission amplitudes at the
two quasienergies are exponentially small in system size $L$, and, in
the limit of large system size, the reflection blocks, $r(0)$ and
$r(\pi)$, become unitary matrices.

\subsection{Topological invariants}

 In five of the ten symmetry
classes, unitarity of the reflection matrix allows us to define topological invariants, along the
lines of the scattering theory of topological insulators and
superconductors.\cite{fulga_scattering_2011}  These classes are AIII,
CII, D, BDI and DIII, as defined in Table \ref{table}, where we also
summarize the main results of this section.

As a first step towards defining the topological invariants, a change
of basis is performed separately for both in- and outgoing lead
states, to simplify Eqs.~\eqref{eq:trs-of-reflection-block},
\eqref{eq:cs-of-reflection-block}, and
\eqref{eq:phs-of-reflection-block}.  Concrete recipes for the basis
transformations are presented in
Appendix~\ref{sec:basis-transformation-appendix} for each class.
{In the thus standardized
  form, the reflection matrices obey the following relations,}
\begin{align}
\label{eq:r_standard_D}
r(\epsilon)&=r^*(-\epsilon)\hspace{59pt} {\rm for\; class\; D},\\
\label{eq:r_standard_DIII}
r(\epsilon)&=r^*(-\epsilon)=-r^T(\epsilon)\hspace{15pt} 
{\rm for\; class\; DIII},\\
\label{eq:r_standard_BDI}
r(\epsilon)&=r^\dag(-\epsilon)\hspace{20pt} 
{\rm for\; classes\; AIII,\; CII\; and\; BDI},
\end{align}
which we 
need to define the topological invariants.
These follow from PHS, PHS + TRS and CS respectively after the simplifying
basis changes.

In class D, $r(0)$ and $r(\pi)$ are real and due to unitarity they are orthogonal matrices.
Hence they have determinant $\pm 1$. Four topologically distinct
situations arise, distinguished by the $\mathbb{Z}_2 \times
\mathbb{Z}_2$ invariant
\begin{align}
	\mathcal{Q}_{\rm D}=\tfrac{1}{2}\text{Det}[r(0)] \times
       \tfrac{1}{2} \text{Det}[r(\pi)] \hspace{10pt} \text{for class D}.
       \label{eq:class-d-invariant}
\end{align}

In symmetry class DIII, the reflection matrices $r(0)$ and $r(\pi)$
are both real and antisymmetric. Therefore, the invariant
of~\eqref{eq:class-d-invariant}, will be $(\tfrac{1}{2},\tfrac{1}{2})$, as the eigenvalues
of real antisymmetric matrices are purely imaginary and come in
complex conjugate pairs.  However, the determinant of an antisymmetric matrix 
is the square of a function of the matrix, the Pfaffian. The Pfaffian in this case can
take values $\pm 1$. Thus, again four topologically different cases
can be distinguished,
\begin{align}
       \mathcal{Q}_{\rm DIII}=\text{Pf}[r(0)] \times
        \text{Pf}[r(\pi)] \hspace{10pt} \text{for class DIII}.
\end{align}

In symmetry classes AIII, BDI, CII the reflection blocks $r(0)$ and
$r(\pi)$ are Hermitian and unitary. Thus their eigenvalues are pinned
to $\pm 1$ and their traces are quantized to integer values. This is
expressed by the $\mathbb{Z}\times \mathbb{Z}$ topological invariant
\begin{align}
\mathcal{Q}_{\rm ch}&= \tfrac{1}{2} \text{Tr}[r(0)] 
\times \tfrac{1}{2} \text{Tr}[r(\pi)] 
\hspace{8pt}\text{for$\,$classes$\,$AIII,CII,BDI}. 
\end{align}
In class CII, the traces can only take even integer values due to
Kramers degeneracy of the scattering states.  In principle, this
invariant is also defined for $i\,r(0)$ and $i \, r(\pi)$ in symmetry
class DIII, which we described before, but will always take the
trivial value $(0,0)$, due to the antisymmetry of $r$.

In combination with the scattering formalism in
Sec. \ref{sect:scattering}, the topological invariants
$\mathcal{Q}_{\rm D}$, $\mathcal{Q}_{\rm DIII}$ and $\mathcal{Q}_{\rm
  ch}$, are the main results of this work.  Our approach is in
agreement with the most recent analysis of topology in DTQWs from a
Floquet operator perspective,\cite{asboth_2013} as we will demonstrate
for three examples in the next section.  Similar invariants exist for
reflection matrices of time-independent systems at zero
energy,\cite{fulga_scattering_2011} but the time-periodicity of DTQWs
leads to an extra contribution at quasienergy $\pi$.

\subsection{Boundary states}

The main reason bulk topological invariants are interesting is
  that they can be used to predict the number of protected midgap
states at an interface between two bulk systems.\cite{rmp_kane} This
 applies to inhomogenous DTQWs that have two
domains, A ($x<0$) and B ($x>0$), governed by different quantum walk
protocols, given that the complete system has the right combination of
symmetries. If the topological invariant
$\mathcal{Q}_X=\mathcal{Q}_{X,0}\times \mathcal{Q}_{X,\pi}$ with $X
\in \{{\rm D,DIII,ch}\}$ changes across the interface by $\Delta
\mathcal{Q}_X=\Delta \mathcal{Q}_{X,0} \times \Delta
\mathcal{Q}_{X,\pi}= \mathcal{Q}_X^A- \mathcal{Q}_X^B$, it can be shown that a number of $|\Delta
\mathcal{Q}_{X,\{0,\pi\}}|$ quasienergy
eigenstates are guaranteed to exist at quasienergies $\varepsilon=0,\pi$ inside the gaps. These are bound to the
interface and protected by the change of topological invariant. A full
discussion based on reflection matrices  is provided in
Appendix~\ref{sec:protected-boundary-states}.

In order to interface two DTQW protocols, such that they form
an inhomogeneous system, the two protocols have to be compatible
(we explain what we mean by this below).  The shift operators
are nonlocal, and thus to ensure that the Floquet operator of the combined system is unitary, 
they have to be applied throughout the system at the same time, and to the same
internal states. Thus, two DTQW protocols A and B are compatible if $S_j^{A} = S_j^{B}$ for every $j$.  The two DTQW protocols can only
differ in their rotations.

Note that there is no unique DTQW analogue of open boundary conditions. Thus 
the bulk topological invariant alone does not predict the number of
topologically protected edge states at the ends of a finite line
segment on which an otherwise homogeneous DTQW takes place.  Edge
states can exist, but their number depends on the way the walk is
terminated.\cite{asboth_prb} 
This is analogous to the situation of time-independent Hamiltonian
systems with chiral symmetry.\cite{fulga_scattering_2011}

Note further that the values of the topological invariants depend on the
starting time of the period of the DTQW, i.e., the choice of time
frame for the Floquet operator.  Nevertheless, the correct number of
protected boundary states is obtained from the individual topological
invariants of two interfaced quantum walk domains when their starting
times are chosen such that the walks are interfacable.

\section{Topological invariant of unbalanced quantum walks}
\label{sect:topology_winding}

When a period of the quantum walk
protocol contains a different number of shift
operators that shift to the right than shift operators
that shift to the left, $n_+ \neq n_-$, the quasienergy bandstructure shows a winding in quasienergy space.\cite{kitagawa_periodic}
This unique type of topology only can occur because of the $2\pi$-periodicity of quasienergy space. From a transport point of
view, such a winding is produced when particles are pumped through the
one dimensional system.
  A simple example is given by
$F=S^+_\uparrow$ for which the quasienergy band structure is given by
the raising half of the green dotted line in
Fig. \ref{fig:standardQW}.

The scattering matrix of such a system has an unusual form since the
reflection blocks $r$ and $r'$ of the scattering matrix are
rectangular matrices of size $n_- \times n_+$ and $n_+ \times n_-$
respectively, while the transmission blocks are square matrices of
differing sizes: $n_+ \times n_+ \;(t)$ and $n_- \times n_- (t')$. The
ranks of the matrix products $r r^\dag$ and $r' r'\vphantom{r}^\dag$
is thus at most as large as $\min (n_+,n_-)$ and one of them has at
least $|n_+-n_-|$ zero eigenvalues. Due to the unitarity of the
scattering matrix, $|n_+-n_-|$ of the transmission eigenvalues of the
larger transmission block have thus to be unity for all
quasienergies. These perfectly transmitting channels in only one direction
reflect the charge pumping through the system. Hence the topology of
the quantum walk can be read off from the scattering matrix through
the topological invariant
\begin{align}
\mathcal{I}=\dim(t)-\dim(t').
\end{align}


\section{Examples}
\label{sec:Examples}

In this section, we consider three examples for gapped DTQWs and
demonstrate how their topological properties can be analysed by the
scattering matrix approach. We first discuss the so-called split-step
walk,\cite{kitagawa_exploring} which includes the simple quantum walk
of Eq.~\eqref{eq:simplQW} as a special case.  We then discuss a
generalization of this protocol, which contains four shift operators
per period.\cite{asboth_2013} Depending on the choice of parameters,
it can fall into several of the relevant symmetry classes, realizing
either $\mathcal{Q}_{\rm D}$ or $\mathcal{Q}_{\rm ch}$. The third
example has a larger internal space and is characterized by the
invariant $\mathcal{Q}_{\rm DIII}$.

Finally, we show that the scattering matrix approach can also be used
to define topological invariants in the presence of disorder and
illustrate this using the simple quantum walk with disordered
rotation angles.

\subsection{Split-step walk}
\label{subsec:splitstep}

Extending the DTQW of Eq.~\eqref{eq:simplQW} by adding another
rotation, we obtain the so-called split-step
walk\cite{kitagawa_introduction}
\begin{align}
\FFF &= S^+_\uparrow R_2 S^-_\downarrow R_1.
\label{eq:split-step-floquet}
\end{align}
Here, $R_j=R(\theta_j)$ is a rotation about the $y$ axis as
  defined in Eq.~\eqref{eq:real-spin-rotation-matrix}. The split-step
  walk is thus parametrized by two angles $\theta_1, \theta_2$. 
This DTQW
has two internal states ($N=2$), again referred to as a spin, with
spin-up propagating to the right, and spin-down
propagating to the left.
Since
  $d_1=d_2=1$, according to 
Sec.~\ref{sect:scattering}, the reflection matrix 
is a $1 \times
1$-matrix.

To find the topological properties of the split-step walk, we first
need to understand its symmetries. According to
Eq.~\eqref{eq:real-spin-rotation-matrix}, the rotation matrices are
real matrices. The same applies for the shift matrices in position
basis, so that $\FFF$ will be real and thus have PHS, with
$\mathcal{P}=K$.\cite{kitagawa_exploring} The protocol also has a
chiral symmetry. This can be seen by choosing a different time frame,
\cite{asboth_2013}
\begin{align}
    \FFF' &=  \sqrt{R_1} S^+_\uparrow R_2 S^-_\downarrow \sqrt{R_1},
\label{eq:split-step-floquet-symmetric}
\end{align}
so that chiral symmetry is given by $\Gamma=\sigma_x$, which can be seen from 
$\sigma_x S_\uparrow \sigma_x= S_\downarrow^{-1}$ and $\sigma_x R \sigma_x = R^{-1}$. Thus the system falls in symmetry class BDI.
Note that also the simple quantum walk is of this form if written as in Eq.~\eqref{eq:simplQW2}, with $\theta_1=0$.

We calculated the reflection matrix in Eq.~\eqref{eq:scat_overlap}
numerically for the Floquet operator $\FFF'$, following the procedure
described in Appendix~\ref{sec:numerical-implementation}.  The
resulting class BDI invariant $\mathcal{Q}_{\rm ch}$ is plotted in
Fig.~\ref{fig:class_bd1} as a function of the rotation angles
$\theta_1, \theta_2$ for system size $L=50$. The calculation is
simplified by the fact that the chiral symmetry of $r$ is in its
canonical form, Eq.~\eqref{eq:r_standard_BDI}, because $ V_\Gamma =
1$. The topological invariant $\mathcal{Q}_{\text{ch}}$ is thus in
fact half of the reflection matrix's only element, taken at energies
$0$ and $\pi$, with values $\mathcal{Q}_{\text{ch}} \in (\pm
\tfrac{1}{2}, \pm \tfrac{1}{2})$.

The results plotted in Fig.~\ref{fig:class_bd1} are in agreement with
topological invariants that were derived directly from the Floquet
operator, by counting gap closings in the dispersion
relation.\cite{asboth_prb}

\subsection{Four-step walk}

We now turn to a
multistep walk, choosing a longer sequence
that includes three different rotations,
\begin{equation}
    \mathcal{F}=S^-_\downarrow R_{3} S^-_\downarrow R_{2} S^+_\uparrow R_{1} S^+_\uparrow.
    \label{eq:twoStepSplitStep}
\end{equation}
Here, we also allow for more general rotations
\begin{align}
R(\theta,\chi)= \sum_x |x\rangle\langle x| \otimes e^{-i\theta (\sigma_y \cos \chi  + \sigma_z \sin \chi)},
\label{eq:parametrization-rotation}
\end{align}
so that the walk is parametrized by six angles, $\theta_j, \chi_j$,
with $ j\in {1,2,3}$.  This four-step walk has been introduced before in
Ref.~\onlinecite{asboth_2013}.

For the four-step walk, there are still only two internal states ($N=2$), but the number of shift operators is larger ($d_\uparrow= d_\downarrow = 2$), leading to a $2 \times 2$ reflection matrix.  The symmetries of the system
are fixed by restricting the parameters to certain subsets. To be
precise, if we set $\chi_{1,2,3}=0$, the rotation matrices are real,
and the system has PHS, given by $\mathcal{P}=K$.  On the other hand,
if we require $R_1=R_3$, the system has chiral symmetry given by
$\Gamma=\sigma_x$.  This
walk thus serves as an illustrative example for the symmetry classes
D, AIII, or BDI.  We concentrate on the BDI case, where all
$\chi=0$ and $\theta_1=\theta_3$.

\begin{figure}
	\centerline{\includegraphics[width=0.8\columnwidth]{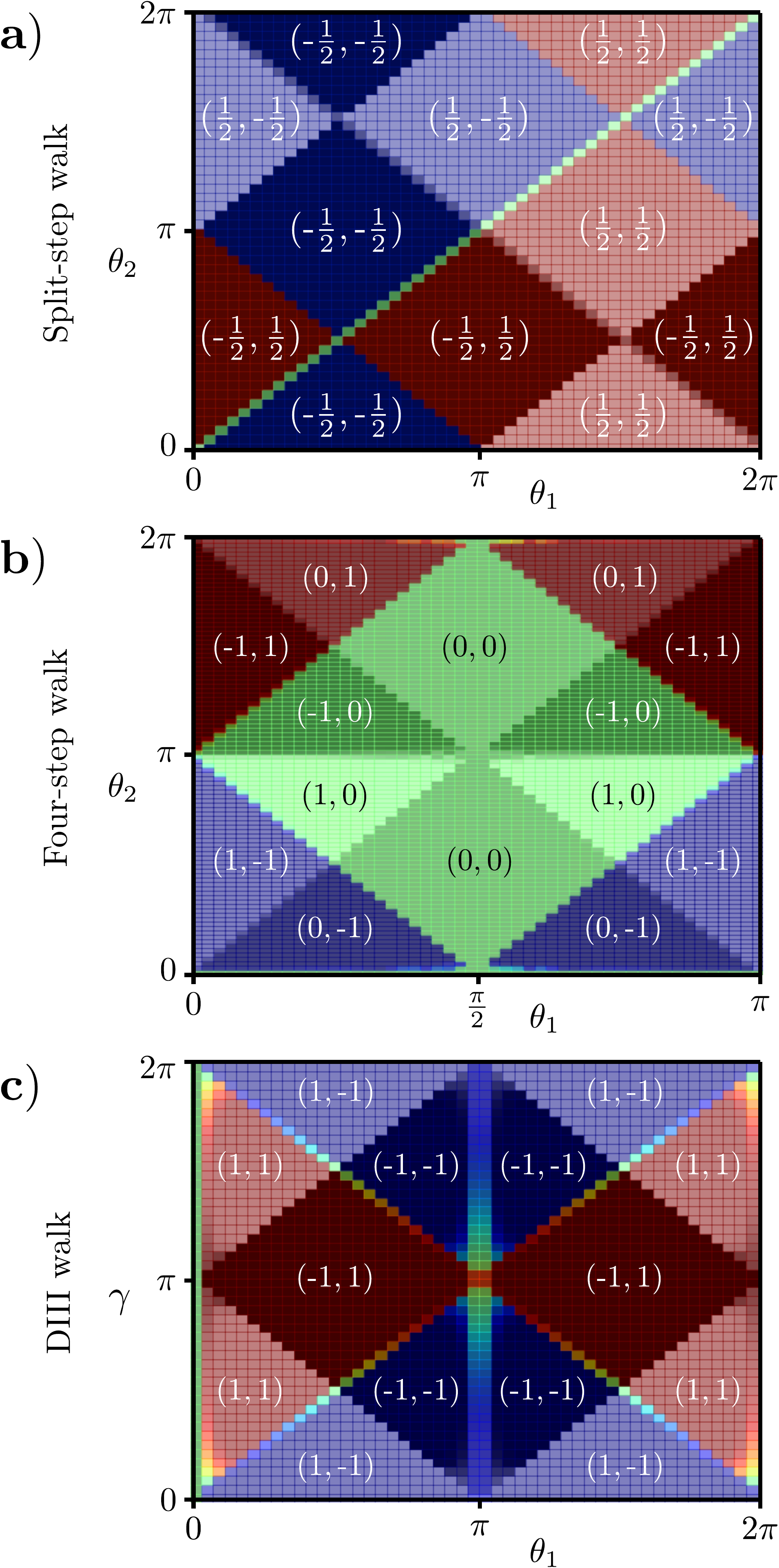}}
    \caption{ Topological phase diagrams for three quantum walk examples, obtained from the scattering matrix approach. All phases are labelled by their topological invariant $\mathcal{Q}_X$ and are furthermore encoded in brightness ($\mathcal{Q}_{\text{X},0}$) and hue ($\mathcal{Q}_{\text{X},\pi}$).     
    {\bf{a)}} Topological invariant
      $\mathcal{Q}_{\text{BDI}}$ of the split-step quantum walk
      \eqref{eq:split-step-floquet-symmetric}.  {\bf{b)}}
      Topological invariant $\mathcal{Q}_{\text{BDI}}$ of the
      four-step quantum walk \eqref{eq:twoStepSplitStep}, where
      $\theta_1=\theta_3$ and $\chi_{1,2,3}=0$, so that falls into class BDI.  {\bf{c)}} Topological invariant
      $\mathcal{Q}_{\text{DIII}}$ of the quantum walk
      \eqref{eq:class-diii-walk}, with $\theta_2=0$.  
      For all three examples, the length of the
      scattering region is $L=50$.  Close to phase
      boundaries, where the gap closes, $r$ becomes subunitary due to
      finite size effects and the invariants are not quantized. Otherwise the quantization of the
      invariants is evident.}
    \label{fig:class_bd1}
\end{figure}

In Fig.~\ref{fig:class_bd1}, we show the numerical result for the
invariant $\mathcal{Q}_{\text{ch}}$ from the scattering matrix.  As
defined in Sec.~\ref{sect:topological_gapped}, the invariant is half
the trace of the reflection block at quasienergies $0$ and $\pi$, and
here each of the two elements can take the values $\{-1,0,1\}$.
Similar to the split-step walk above, the symmetry relations for $r$
are in their standard form already, so no basis transformation is
required.

Our result for the the phase diagram agrees with Fig. 2 of
Ref.~\onlinecite{asboth_2013}, where the topological invariant was
calculated by combining winding numbers from two different time
frames. Interestingly, with the approach of this paper, it suffices to
consider the protocol in one time frame. This is because the
scattering matrix method uses all possible plane waves to probe the
quantum walk, which reach the scattering region at different
times. The reflection matrix thus contains information about the
dynamics of the system during one timestep.

Quantum walks for classes AIII and D are obtained from this walk by
breaking either particle-hole or chiral symmetry. In the former case,
the topological invariant does not change, while in the latter case,
the topological invariant is reduced to
$\mathbb{Z}_2\times\mathbb{Z}_2$. \cite{asboth_2013}

\subsection{Symmetry class DIII}

The construction of a DTQW that realizes  $\mathcal{T}^2=-1$ is more involved;
some proposals  have been given in Ref.~\onlinecite{kitagawa_exploring}.
As an example, we now consider a protocol with DIII symmetry, which is constructed with four internal states $N=4$,
of which two are right-moving and two are left-moving. We consider these as two instances of a two-state quantum walk,
 which are governed by
\begin{equation}
    \FFF =
	\begin{pmatrix} 
		F_1 & 0 \\
		0   & F_2
	\end{pmatrix}
	e^{i \sigma_z \tau_y \gamma}
	\begin{pmatrix} 
		F_2 & 0 \\
		0   & F_1
	\end{pmatrix},
    \label{eq:class-diii-walk}
\end{equation}
where $\sigma_i$ are Pauli matrices acting on the spin of each copy of the
two-state quantum walk, while $\tau_y$ is a Pauli matrix that mixes the two instances. Here, $F_{1}$ and $F_{2}$  are both Floquet operators of the simple quantum
walk in the form of Eq.~\eqref{eq:simplQW2}, with different parameters
$\theta_{1/2}$. The additional angle $\gamma$ provides a way to couples the two instances of the
walk.  This quantum walk has CS with $\Gamma=i\sigma_x\tau_y$, PHS with
$\mathcal{P}=K$, and thus TRS with $\mathcal{T}=\sigma_x\tau_y K$, falling
into symmetry class DIII.

According to section \ref{sect:topological_gapped}, the calculation of
the topological invariant from the reflection block $r$ requires us to
find the basis in which $r$ is antisymmetric, in order to
calculate the Pfaffian. From
Appendix~\ref{sec:basis-transformation-appendix}, it follows that this
property is fulfilled by the matrix $\tilde{r}=V_T r$, so that the
topological invariant in this example can be calculated as
\begin{equation}
    \mathcal{Q}_{\text{DIII}} = \text{Pf}\left( \tau_y r(0) \right)  \times \text{Pf}\left( \tau_y r\left( \pi \right) \right) .
    \label{eq:PfaffianForExampleWalk}
\end{equation}

The resulting phase diagram of this protocol, with $\theta_2=0$, is
displayed in Fig.~\ref{fig:class_bd1} c). It realizes all possible
topological phases of the symmetry class. Non-generic features can be observed at
$\theta_1=0,\pi$ in the phase diagram, signalling unprotected gap closings at which the topological invariant does not change.

\subsection{Disorder}

A major advantage of the classification of
topological phases using the scattering matrix is that the 
topological invariants can also be defined for systems with spatial disorder.

As a proof of concept, let us now add disorder to the simple quantum walk,
Eq.~\eqref{eq:simplQW2}.  Spatial disorder is introduced by
drawing the the rotation angle $\theta\left( x \right)$ for each site $x$ from a Gaussian
ensemble with mean $\left<\theta\right>$ and variance $\delta\theta$, with no correlation for different $x$.
This breaks neither PHS nor CS, so a BDI topological invariant is still defined if $r$ remains unitary.

\begin{figure}
	\includegraphics[width=0.8\linewidth]{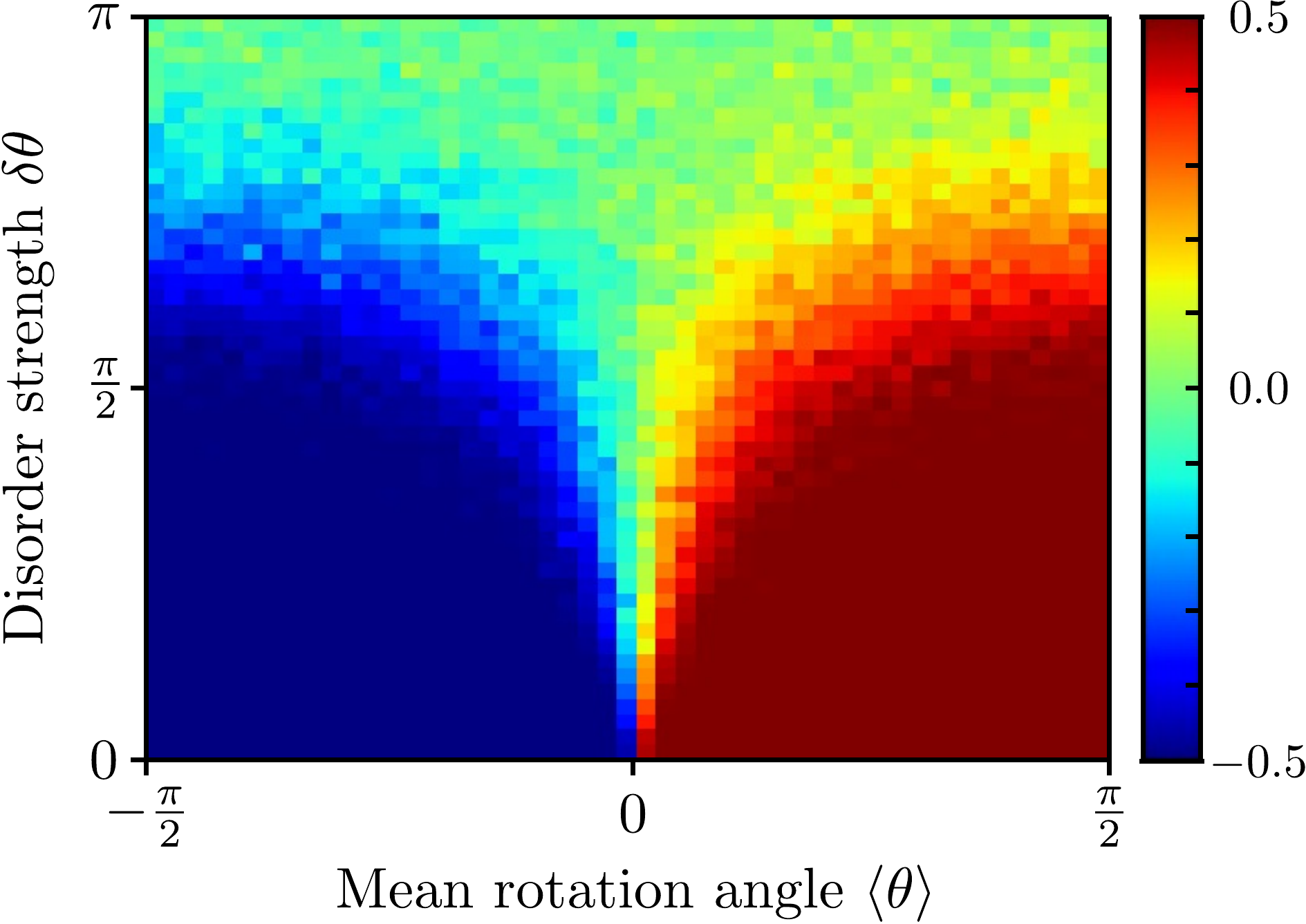}
    \caption{Disorder averaged invariant $\langle Q_{{\rm ch},0} \rangle$ at $\varepsilon=0$ for the simple quantum
        walk as a function of mean rotation angle and disorder strength. The transition region around $\langle \theta \rangle=0$ is broadened with increasing disorder 
        until the topological phases are not properly defined anymore (green region). 
    System size is $L=50$, the average is taken over $n=100$ different disorder realizations.}
    \label{fig:disorder}
\end{figure}

As for the split-step walk, the BDI topological invariant is just half the reflection block itself, which is a single number. 
Furthermore, due to an additional symmetry,\cite{kitagawa_introduction} $\mathcal{Q}_{{\rm ch},\pi}=-\mathcal{Q}_{{\rm ch},0}$, so we only have to consider quasienergy $\epsilon=0$. 
We thus numerically calculated an ensemble average of $r(0)$ for a range of $\left<\theta\right>$ and $\delta\theta$ which is presented in Fig.~\ref{fig:disorder}.
Note that the topological invariant is stable against the introduction of small disorder unless very close to the transition, demonstrating the stability of the phases to disorder.

For strong disorder, the ensemble average approaches zero (the green region in Fig.~\ref{fig:disorder}). However, this is not due to the fact that $r$ becomes subunitary. On the contrary,
the distribution of $r$ is strongly bimodal around $\pm 1$, indicating that individual systems are still insulating and allow for the definition of a topological invariant, whose value however can not be predicted 
for large disorder strengths.

\section{Experiment}
\label{sect:experiment} 

The scattering matrix of a discrete-time quantum walk is not only a
theoretical construct but can also be directly measured.  In this
section we discuss the principle of such an experiment using the
example of the split-step walk, introduced in Section
\ref{subsec:splitstep}.  For the split-step walk, the reflection
matrices are real numbers of unit magnitude, $(r(0),r(\pi))=(\pm 1,\pm
1)$, and, using \eqref{eq:scat_overlap}, the pair of topological
invariants simplify to
\begin{align}
\mathcal{Q}_{\rm ch,0} &= \frac{1}{2}\sum_{\nu=1}^{\infty} 
\bra{-1,\downarrow} \FFF^\nu \ket{-1,\uparrow}; \nonumber \\
\mathcal{Q}_{\rm ch,\pi} &= \frac{1}{2}\sum_{\nu=1}^{\infty} (-1)^\nu
\bra{-1,\downarrow} \FFF^\nu \ket{-1,\uparrow}.
\label{eq:splitstep_q}
\end{align}
These formulas suggest a measurement protocol for the topological
invariants:  1)
Initialize the walk with the walker at time $\tau=0$ at $x=-1$, in
state $\uparrow$. 2) Obtain the topological invariants as the sum, and
alternating sum of the probability amplitudes for the walker at
timestep $\tau\in\mathbb{N}$ to be at $x=-1$, in state
$\downarrow$. 
This measurement can be 
straightforwardly
conceived in optical realizations of quantum walks, as we show
below. 
 

\begin{figure}
  \includegraphics[width=0.8\columnwidth]{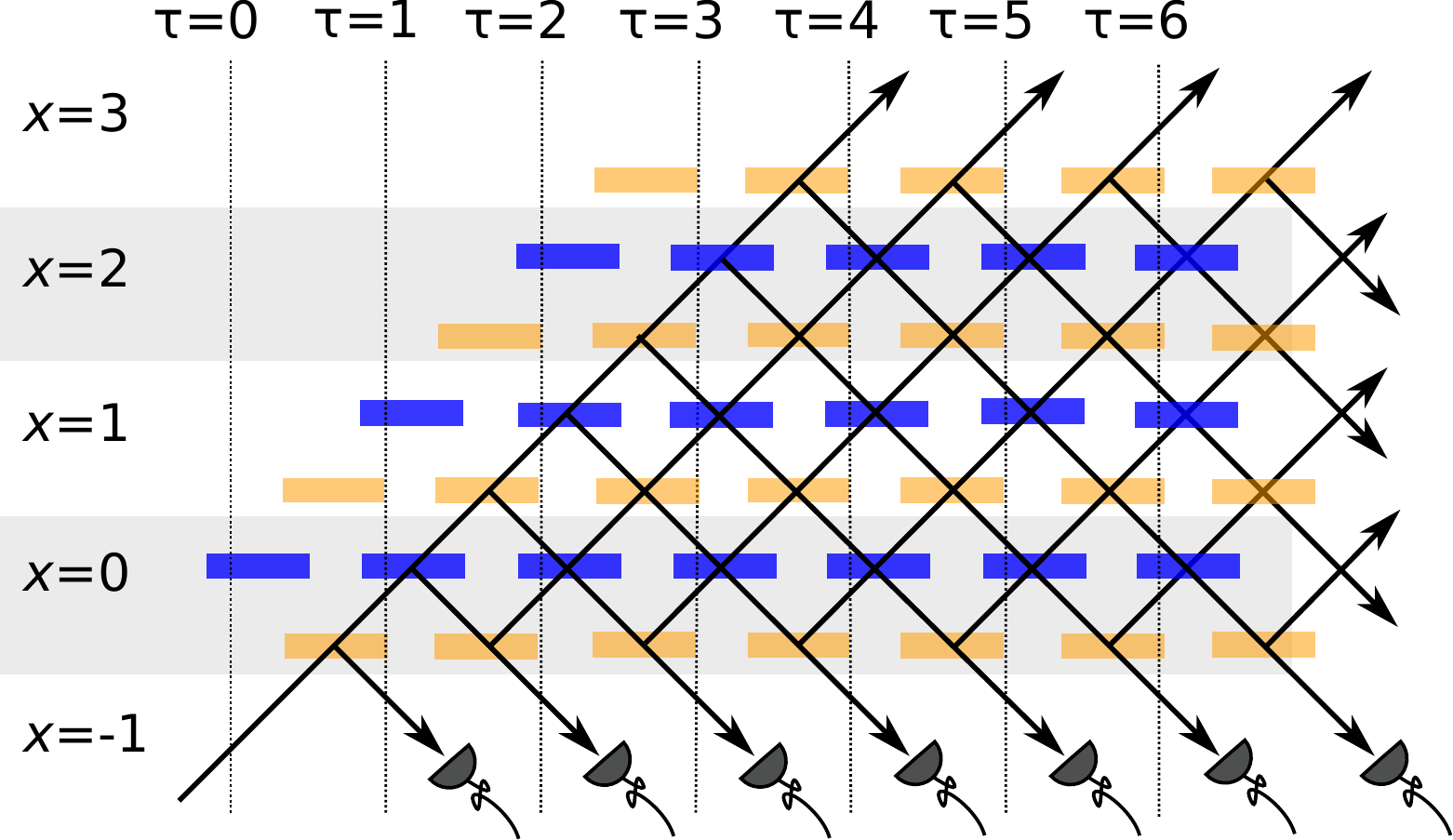}
  \caption{Schematic layout for the experimental measurement of the
    reflection amplitudes of a split-step quantum walk.  An incident
    coherent light pulse at $\tau=0$, $x=-1$ enters an array of beam
    splitters of two types (dark blue, light orange), where it is
    split and recombined repeatedly, thereby performing the quantum
    walk. A row of detectors at $x=-1$ measure the wave amplitudes
    $\bra{-1,\downarrow} \FFF^\tau \ket{-1,\uparrow}$ leaving the
    quantum walk region.  The reflection amplitudes $r(0)$ and
    $r(\pi)$ are given by the sum and and the alternating sum of the
    measured reflected amplitudes, Eqs.~\eqref{eq:splitstep_q}.}
  \label{fig:experiment2}
\end{figure}



We demonstrate our ideas using a simple beam splitter (BS)
representation of the quantum walk, shown in
Fig.~\ref{fig:experiment2}. This layout can be easily adapted to many
actual physical realizations, including integrated
photonics,\cite{sciarrino_twoparticle} or even optical feedback
loops. \cite{schreiber_science}  It consists of an array of cascaded
BS's, with a light pulse incident on the lower left BS.  As the light
propagates in time, it spreads throughout the array in a way that can
be interpreted as a quantum walk. The state of the light just before
and just after the $n$th column of BS's is mapped to the state of the
walker just before and just after the $n$th rotation operation. The
direction of propagation of the modes is identified with the internal
state of the walker, ``right-up'' representing $\uparrow$ and
``right-down'' representing $\downarrow$. The vertical coordinate in
the arrays is identified with the position $x$ of the walker, as
indicated in Fig.~\ref{fig:experiment2}.  We use two different types
of BS's to realize the two rotations in the Floquet operator,
Eq.~\eqref{eq:split-step-floquet}.

In optical DTQW experiments, intensity measurements on the modes
leaving the array at the right edge are used to read out the
position distribution of the walker after $\tau$ steps. In our case,
there are two differences. First, as indicated 
in Fig.~\ref{fig:experiment2}, our output modes are not at the right edge, but rather
at the bottom edge of the array. Second, intensity measurement on the
output modes does not work for us, since it destroys the phase
information that is crucial to obtain the topological invariants, 
as sums of probability amplitudes, Eq.~\eqref{eq:splitstep_q}. 

A direct measurement of the probability amplitudes as required for the
invariants is possible if the incident light pulse is a strong
coherent state $\ket{\alpha}$, containing many photons. 
This is
standard practice in some photonic quantum walk
experiments.\cite{schreiber_science}
Strictly speaking the spreading of the light pulse is then
not a quantum walk any more, since there is no entanglement at any
point in the system. However, it simulates a single-photon quantum
walk directly. At any time, the array contains coherent states
$\Pi_j \ket{\alpha_j}$, with the coherent amplitudes $\alpha_j$
corresponding exactly to the probability amplitudes $\Psi_j$ of the
walker, $\Psi_j = \alpha_j/\alpha$. This is used in
experiments\cite{gabris_prl} to read out the state of the walker
during the walk, and to measure the probability distribution after $N$
steps in one shot.

The $0$ and $\pi$ quasienergy invariants are obtained by measuring the
sum and the alternating sum of the outcoming coherent amplitudes,
cf. Eqs.~\eqref{eq:splitstep_q}. This can be done practically by
interfering each output mode with a local oscillator, or, interfering
the output modes directly with each other on an N-port. Note that
since the BS's have only real elements (no phase shifting), a single
intensity measurement suffices. Moreover, in this setup, one can even
use a CW laser instead of a laser pulse.

\section{Conclusion}

In this paper we have classified the topological phases
of one-dimensional discrete-time quantum walks using a scattering
matrix approach. For this purpose, we generalised the concept of the
scattering matrix to these periodically time-dependent systems. 

We find that, dependent on their symmetries, gapped DTQWs are
characterised by one of three different topological
invariants, $\mathcal{Q}_{\rm D}$, $\mathcal{Q}_{\rm DIII}$ and
$\mathcal{Q}_{\rm ch}$. They are calculated from the determinant,
Pfaffian or trace of the reflection matrix as summarised in Table
\ref{table}. In contrast to their analogs for time-independent
systems,\cite{fulga_scattering_2011} the invariants consist of two
independent contributions $\mathcal{Q}=\mathcal{Q}_0 \times
\mathcal{Q}_\pi$ that are evaluated at the two special quasienergies
$\varepsilon=0,\pi$.  Adapting arguments for topological
insulators,\cite{fulga_scattering_2011} we found that an interface
between two extended quantum walk regions hosts a number of protected
boundary states that equals the difference of the invariants across
the interface. These are stationary states of the walk where the
walker stays exponentially close to the interface, and has quasienergy
$\varepsilon=0$ or $\varepsilon=\pi$.

We also considered unbalanced DTQWs where there is a difference $n$
in the number of left- and rightward shifts per cycle, producing a
quasienergy winding in the Brillouin zone. We found that they have $n$
channels that transmit perfectly in the majority direction. The
characterization of transmission in this problem, including the
transport time distribution of disordered quantum walks with
quasienergy winding, poses an interesting direction for further
investigation.

We provide a simple scheme to directly measure the reflection matrix
-- and, thus, the topological invariants -- of a quantum walk. This
scheme is well within the reach of current experiments working with
light pulses \cite{peruzzo_science_2010, sciarrino_twoparticle,
  schreiber_science, gabris_prl}.
  
Our scattering matrix approach complements existing methods based on
Floquet operators in momentum space, with two important
advantages. First, we provide a unified framework describing
topological phases in different symmetry classes as simple functions
of a single, typically small matrix. Second, 
our formulas use only a single time frame for the
Floquet operator.
This is in contrast with Ref. \onlinecite{asboth_2013}, which
explicitly states that the topological invariants of chiral quantum
walks can only be obtained by combining the winding numbers from
different timeframes. The scattering matrix gets around
  this restriction, and probes the behaviour of the system
  \emph{during} a protocol by including contributions from plane
  wave-like modes that enter and exit the scattering region at
  intermediate times.

The scattering matrix formalism introduced in this paper gives a
powerful new tool for the investigation of the effects of disorder on
topological phases and transport in DTQWs. Depending on the types of
disorder and symmetries, experiments and theory on DTQWs have already
seen both Anderson localization, \cite{gabris_anderson} and
delocalization. \cite{obuse_delocalization} Our generalized scattering
matrix formalism allows a continuation of this research to more
general multistep DTQWs.

\acknowledgements

We thank Carlo Beenakker for his input and support.  JKA thanks
A.~G\'abris for helpful discussions.  This research was realized in
the frames of TAMOP 4.2.4. A/1-11-1-2012-0001 ''National Excellence
Program -- Elaborating and operating an inland student and researcher
personal support system'', subsidized by the European Union and
co-financed by the European Social Fund.  This work was also supported
by the Hungarian National Office for Research and Technology under the
contract ERC\_HU\_09 OPTOMECH and the Hungarian Academy of Sciences
(Lend\"ulet Program, LP2011-016).  The work was further supported by the
Foundation for Fundamental Research on Matter (FOM), the Netherlands
Organization for Scientific Research (NWO/OCW), an ERC Synergy Grant,
the EU Network NanoCTM, and the German Academic Exchange Service
(DAAD).

\appendix

\section{Numerical implementation}
\label{sec:numerical-implementation}

According to Eq.~\eqref{eq:scat_overlap} the scattering matrix is
determined by following the time evolution of a particle which is placed
in an incoming mode until it enters an outgoing mode.
While doing so, most of the infinite Hilbert space of the scattering
problem will not be reached by the particle. Consequently, we can
evaluate this formula in a modified, finite Hilbert space.

We thus introduce a reduced circular system, which contains all states of the
$L$ sites in the system, and additional ``buffer'' states, which we now describe.
Consider all lead states that are localised on a single lead site only and,  after one period, will be shifted into the scattering region. These are the only lead
states which are non-trivially involved during one time step, all other lead states are just shifted according to the lead propagator.
Likewise, consider all localised lead states that are reached from the scattering region during one period. 
These two groups of states are arranged symmetrically with respect to the
scattering center: Whenever a shift operator moves a state into  the scattering region from one side, a corresponding state on the other side of the system is moved out of the system.
To form the reduced finite space, we identify such pairs of lead states with each other. Each pair forms one of the buffer states, which in turn
form a circular system when combined with the scattering region.

In summary, there are $d_n$ buffer states and $L$ system states
in the reduced space for each internal state $n$.
For exactly one time period, the time evolution of this finite system will be the same as for the original infinite system. 

We can use this system to describe the complete scattering process, if before each step we
initialize the buffer states with a wave function from the incoming leads,
propagate for one unit of time, and then unload the buffer states as the
outgoing mode. Denoting by $\psi_\text{sys}$ the wave function on the scattering sites and by $\psi_\text{in/out}$ the states of the buffer, the dynamics are described by:
\begin{equation}
    \begin{pmatrix}
        \psi_{\rm sys}\left( t+1 \right)\\
        \psi_{\rm out}\left( t+1 \right)
    \end{pmatrix}
    =
    V
    \begin{pmatrix}
        \psi_{\rm sys}\left( t \right)\\
        \psi_{\rm in}\left( t \right)
    \end{pmatrix}
    =
    \begin{pmatrix}
        A & w_\text{in} \\
        w_\text{out} & S_0
    \end{pmatrix}
    \begin{pmatrix}
        \psi_{\rm sys}\left( t \right)\\
        \psi_{\rm in}\left( t \right)
    \end{pmatrix},
    \label{eq:fyod-summ-formula}
\end{equation}
where the matrix $V$ describes the effect of $\mathcal{F}$ on this
reduced space.  We note that this form corresponds to the standard
form for discrete-time scattering problems given in
Ref.~\onlinecite{fyodorov_2000}.

We can write $V$ in terms of modified shift and rotation operators:
\begin{equation}
    V = V_S^{(M)} V_R^{(M)}\cdots V_S^{(2)}V_R^{(2)}V_S^{(1)}V_R^{(1)}.
    \label{eq:V-from-F}
\end{equation}
Here, the effect of $S^{\left( j \right)}$ on our reduced space is given by
a shift matrix
\begin{align}
  V_{S}^{\left( j \right)} &=\sum_n \sum_{x=-d_n}^{L}
    \ket{x+s_{n_j},n}\bra{x,n},
\label{eq:definition-vs}
\end{align}
which is circular because of the identification of incoming and outgoing localized states $\ket{L+1,n}\simeq\ket{-d_n,n}$.
Similarly, the effect of a rotation on this space is given by
\begin{eqnarray}
    \lefteqn{V_R = \sum_{n,n'}\sum_{x=1}^{L} \ket{x,n}R_{nn'}(x)\bra{x,n'}} \nonumber\\
    & &  {} + \sum_{n}\sum_{x=-d_n}^{0} \ket{x,n}\bra{x,n},
    \label{eq:definition-vr}
\end{eqnarray}
applying the rotation to the system, but not to the buffer.

It can then be shown\cite{fyodorov_2000} that the scattering matrix (reflection and transmission) can be obtained from the finite matrix $V$ by
\begin{align}
    S & =  w_\text{out} \left( e^{-i \varepsilon} - A \right)^{-1} w_\text{in} + S_0,
    \label{eq:fyod-summ-scattering-mat}
\end{align}
in contrast to Eq. (\ref{eq:scat_overlap}) which is defined on an infinite space.

\section{Symmetries of the reflection matrix}
\label{sec:basis-transformation-appendix}

\subsection{Derivation of the symmetry relations}

We demonstrate how we obtain the symmetry relations
Eqs.~\eqref{eq:trs-of-reflection-block},
\eqref{eq:cs-of-reflection-block},
\eqref{eq:phs-action-on-lead-states} for the reflection matrix.
Assume that we are given a scattering state with
one incoming mode ($n\in M_+$), so that
\begin{align}
    \left( \varepsilon - H_\text{eff} \right) \left[ \ket{l_{n,d,\varepsilon}} + r \ket{l_{n,d,\varepsilon}}  + \ket{\Psi_C}\right]&=0.
    \label{eq:scattering-state}
\end{align}
The first term is the incoming mode and the second term describes the
corresponding reflected modes, where we use operator notation for the reflection matrix:
\begin{align}
    r\ket{l_{n,d,\varepsilon}} &= \sum_{n'\in M_-} \sum_{d'=1}^{d_{n'}} r_{n'd',nd} \ket{l_{n',d',\varepsilon}}.
    \label{eq:explain-operator-notation}
\end{align}
The third term describes the wavefunction within the scatterer,
cf. Eq.~\eqref{eq:scattering-state-wave-function}.  

By application of
the TRS operator $\mathcal{T}$ on Eq.~\eqref{eq:scattering-state}, using the fact that it commutes with $H_\text{eff}$,
and employing the representation of TRS on the scattering states, Eq.~\eqref{eq:action-of-trs-on-scattering}, we
find that
\begin{align}
    \left( \varepsilon - H_\text{eff} \right) \left[ Q_T \ket{l_{n,d,\varepsilon}} + V_T r^* \ket{l_{n,d,\varepsilon}}  + 
    \mathcal{T}\ket{\Psi_C}\right]&=0,
    \label{eq:trs-scattering-state}
\end{align}
where the complex conjugation occurs due to the antiunitarity of $\mathcal{T}$.

Thus we constructed another scattering state at energy
$\varepsilon$, where the incoming modes are the time-reversed former outgoing modes: $V_T r\left( \varepsilon \right)^*
\ket{l_{n,d,\varepsilon}}$, and outgoing modes are constructed from the
time-reversed incoming mode: $Q_T\ket{l_{n,d,\varepsilon}}$. By the definition
of $r$, we thus must have the relation 
\begin{align}
    r\left( \varepsilon \right) V_T r\left( \varepsilon \right)^* \ket{l_{n,d,\varepsilon}} &= Q_T \ket{l_{n,d,\varepsilon}},
    \label{eq:equation-for-s-from-trs}
\end{align}
and as this holds for all $n\in M_+$ and corresponding $d$, we can conclude Eq.~\eqref{eq:trs-of-reflection-block}. 
Analogous arguments can be given to show Eq.~\eqref{eq:cs-of-reflection-block} and Eq.~\eqref{eq:phs-of-reflection-block}.

\subsection{Basis transformations}

We next consider basis transformations of the incoming and outgoing modes in order to 
turn the symmetries of $r$ presented in Eqs.~\eqref{eq:trs-of-reflection-block} to \eqref{eq:phs-of-reflection-block} into standard form. 
Because the incoming and outgoing modes are separate spaces, we can choose basis transformations for both independently. This amounts to a multiplication of $r$ with two unrelated unitary matrices from the left and right respectively.

In the following we assume that $r$ is taken at energies $\varepsilon = 0, \pi$ and we suppress energy dependence.

\paragraph{Class D}
If $\mathcal{P}^2=1$, it can be seen that $Q_{P\pm} = Q_{P\pm}^T$. Thus, 
we can find square roots $M^2_{\pm}=Q_{P,\pm}$, which are also symmetric. It can then be checked that after the transformation
\begin{align}
    \tilde{r} &= M_-^* r M_+^T,
    \label{eq:phs-basis-transform}
\end{align}
Eq.~\eqref{eq:phs-of-reflection-block} is equivalent to $\tilde{r}=\tilde{r}^*$. 

\paragraph{Class DIII}
If $\mathcal{T}^2=-1$, one can see that $Q_T^T=-V_T$. Again, we can find symmetric square roots, 
\begin{align}
    M_+^2 &= Q_{P,+}, \\
    M_-^2 &= Q_T^\dag Q_{P,-} Q_T^*,
    \label{eq:basis-trans-matrices-diii}
\end {align}
and performing the basis transformation
\begin{align}
    \tilde{r} &= M_-^* V_T^* r M_+^T,
    \label{eq:basis-transform-diii}
\end{align}
 this leads from
Eqs.~\eqref{eq:trs-of-reflection-block} and \eqref{eq:phs-of-reflection-block}
to $\tilde{r} = \tilde{r}^* = - \tilde{r}^T$. Importantly, one
uses the fact that because of assumed irreducibility of any unitary symmetry operator,
by Schur's lemma we must have $\mathcal{PTPT}=e^{i \phi}$, from which one finds
that $M_-M_+^T=e^{-i\phi/2}$.

\paragraph{Chiral classes}
For these classes, we have a chiral operator, obeying $V_\Gamma Q_\Gamma = \Gamma^2=1$.
Then we can choose $\tilde{r}=V_\Gamma r$ and from Eq.~\eqref{eq:cs-of-reflection-block} find
$\tilde{r}=\tilde{r}^\dag$.

We note that these transformation are not unique (for instance, in class D, any orthogonal transformation preserves $\tilde{r}=\tilde{r}^*$)
, so that other possible choices exist. The actual value of topological invariants
obtained from $\tilde{r}$ depend on the
choice. However, because there is no unambigious notion of a trivial vacuum for quantum walk systems, we do not 
impose further restrictions on the choice of basis, and instead remark that the definition of topological invariants 
is only possible after fixing a specific suitible basis.

\section{Protected boundary states}
\label{sec:protected-boundary-states}

Here we derive the existence of protected boundary states caused by a change of
topology across an interface between two domains with a different DTQW
protocol. We exemplify the derivation for a class D quantum walk.
For other symmetry classes, one can argue in a similar
fashion.~\cite{fulga_scattering_2011} 

If two compatible DTQWs, a left (A) and right (B) one, are
interfaced, a bound state occurs at the interface whenever $\det(1-r_A
r^{\prime}_B)=0$, where $r'$ denotes the reflection matrix for incoming states from the right. 
Consider a fixed energy $\varepsilon \in 0,\pi$. The reflection matrices $r_A$ and
$r'_B$ are orthogonal matrices at this energy, as is their product. Thus,
$\det{(r_A r^{\prime}_B)}=\pm1$. 

The determinant $\det{(r_A r^{\prime}_B)}$ is the product of the eigenvalues of an orthogonal matrix,
which in term come either in complex conjugate pairs or are $1$ or $-1$. 

For
even matrix size and $\det{(r_A r^{\prime}_B)}=-1$, an odd number of eigenvalues has to be
$-1$ and thus at least one eigenvalue $1$.  An eigenvalue of $1$ amounts to a
bound state at the given energy.  For odd matrix size on the other hand,  a
positive determinant requires at least one eigenvalue $1$ and thereby ensures a
bound state.

To connect these bound states to  the topological invariant $\mathcal{Q}_\text{D}$, we first need to understand the relation between $r$ and $r'$. This can be deduced by
requiring that by
connecting two copies of the same quantum walk, no bound states should exist (they
would be states in the middle of a gap). Thus for even matrix dimension, $\det
r = \det r'$ while for odd matrix dimension $\det r = -\det r'$. 

In conclusion
this means that, when $\det r_A \neq \det r_B$, a bound state between the two
regions is ensured by the change of topology across the boundary.

\bibliography{walkbib}
\bibliographystyle{apsrev}

\end{document}